\documentclass[12pt,preprint]{aastex}

\usepackage{natbib}
\usepackage{times}
\usepackage{sidecap}

 \newcommand{\sgn}{\mathop{\mathrm{sgn}}}

\shortauthors{Lionello et al.}

\begin{document}


\title{Application of a Solar Wind Model Driven by Turbulence Dissipation to a
2D Magnetic Field Configuration}
\author{Roberto Lionello}
\affil{Predictive Science, Inc.,  9990 Mesa Rim Rd., Ste. 170, San Diego, CA
92121-3933}
\email{lionel@predsci.com}
\author{Marco Velli} 
\affil{Department of Earth, Planetary, and Space Sciences, UCLA,
595 Charles E.\ Young Drive, East, Los Angeles, CA 90095-1567}
\email{mvelli@ucla.edu}
\author{Cooper Downs, Jon  A.\ Linker, 
Zoran Miki\'c}
\affil{Predictive Science, Inc.,  9990 Mesa Rim Rd., Ste. 170, San Diego, CA
92121-3933}
\email{\{cdowns,linker,mikic\}@predsci.com}


\begin{abstract}
Although it is widely accepted that photospheric motions provide the energy source and that
the magnetic field must play a key role in the process, the detailed 
mechanisms responsible for heating the Sun's corona and
accelerating the solar wind are still not fully understood.
\citet{2007ApJS..171..520C} developed a sophisticated, 1D,
time-steady model of
the solar wind with turbulence dissipation. By varying the coronal magnetic
field, they obtain, for a single choice
of wave properties, a realistic range of slow and fast wind conditions with
a sharp latitudinal transition between the two streams.
Using a 1D, time-dependent model of the solar wind of
\citet{2014ApJ...784..120L}, which incorporates turbulent dissipation of
Alfv\'en waves  to provide
heating and acceleration of the plasma, we have explored a similar
configuration, obtaining qualitatively equivalent results. 
However, our calculations suggest that the rapid transition between slow and fast wind suggested
by this 1D model may be disrupted in multidimensional MHD simulations by the requirement
of transverse force balance.

\end{abstract}
\keywords{MHD --- (Sun:) solar wind --- turbulence -- waves}


\section{INTRODUCTION}
{
Although the physical processes responsible for heating the solar corona
and accelerating the solar wind have not been unambiguously 
identified, it is thought that  the interaction of the magnetic field with 
random motions on the photospheric surface may play a crucial role.
Since measurements of wind speeds in fast streams indicate
that the deposition of heat occurs over extended length scales
\citep{1977ARA&A..15..363W,1980JGR....85.4665H,1988ApJ...325..442W},
 one-dimensional (1D) models of the solar wind in the past have included 
parametric
heating forms involving exponential functions decaying with
the distance above the Sun's surface  \citep{1982ApJ...259..779H,1982ApJ...259..767H,1988ApJ...325..442W,
1995JGR...10021577H,1995GeoRL..22.1465R,1997ApJ...482..498H}. 
More sophisticated treatments have derived acceleration and heating rates for the solar wind 
based on the dissipation of  long-period, large-scale, broadband fluctuations 
\citep{1968ApJ...153..371C,1971JGR....76.3534B,1986JGR....91.4111H,
1988JGR....93.9547H,1994AdSpR..14..123V,1999ApJ...523L..93M,2007ApJ...662..669V,2012ApJ...745...35Z}.
}

In the context of MHD models of the solar coronal and inner heliosphere, 
 because of the very large temporal and spatial ranges that are involved in the dynamics,
it is complicated to connect the large-scale heating formulation for the heating of the plasma and the
acceleration of the wind with the underlying physical
mechanisms.
During the past years and
with different levels of self-consistency,
mechanisms for turbulent dissipation have been integrated into 1D models of the solar wind
\citep{2005ApJ...632L..49S,2005ApJS..156..265C,2007ApJS..171..520C,2010ApJ...710..676C,
2010ApJ...708L.116V,2011ApJ...743..197C,2014ApJ...784..120L}. 
Formulas for turbulence dissipation are also increasingly replacing empirical
heating functions 
in three-dimensional (3D) MHD models such as those of \citet{2009ApJ...690..902L} or \citet{2010ApJ...712.1219D}.
Since these 3D models would have to resolve both the temporal scales related to dissipation in the solar corona
(of the order of milliseconds) and those related to large structures in the solar wind (many days), this integration
 represents
a formidable challenge. Proton heating through Kolmogorov dissipation in regions of open magnetic field, together with
Alfv\'en waves acceleration of the solar wind, was implemented in the model of
\citet{2010ApJ...725.1373V}. The large-scale MHD model of the heliosphere of 
\citet{2011ApJ...727...84U}  includes, for the region where the
solar wind is superalfv\'enic and supersonic, a formulation for the transport and dissipation of turbulence energy.
A phenomenological description of nonlinear interactions, which are ascribed to
wave reflection caused by chromospheric and coronal density gradients, is present in the model of \citet{2013AIPC.1539...30L},
including  the   outward propagation Alfv\'enic turbulence in the solar wind.
\citet{2013ApJ...764...23S} implemented an Alfv\'enic turbulent transport and dissipation mechanism, which both heats
 the coronal plasma and accelerates the solar wind 
{\citep[further details are presented in][]{2014ApJ...782...81V}. } 
Finally, \citet{2012ApJ...749....8M,2014MNRAS.440..971M} proposed
{an integrated, 2.5D model that connects the photosphere and the inner heliosphere.
The model inputs only torsional Alfv\'en waves at the coronal base, allowing coupling with both 
parallel (sound) and transverse magnetoacoustic modes. Reflection is self-consistently included. 
Though both polarizations transverse to the radial are included, there is only one transverse 
direction so that a purely incompressible transverse cascade in the spirit of Reduced Magnetohydrodynamics
 (RMHD) does not occur. Rather Matsumoto and Suzuki see complex compressive interactions between the 
average radial wind and the fluctuations, including parametric decay.
In the same spirit of the works of Matsumoto and Suzuki are the 2D simulations
 of \citet{1998JGR...10323677O} and \citet{2005SSRv..120...67O}.  On the other hand,
it is the features of such a cascade that our simpler 1D models attempt to capture 
phenomenologically \citep{2014ApJ...784..120L}.
}

Notwithstanding the appeal of multidimensional models, 1D simulations
still play a crucial role, since they can reach a level of sophistication
in the physical model and of resolution in the calculation that are not
otherwise achievable. They also constitute fundamental tests to which
the results of multidimensional models must be reconciled.
However, there is also a risk of extending 
the validity of results obtained through 1D simulations
to more realistic configurations in which 
they may not necessarily apply.

\citet{2007ApJS..171..520C} developed a sophisticated, 1D, time-steady
model of
the solar wind, from the photosphere to the inner heliosphere. Their model
uses  sound and Alfv\'en waves for flow acceleration           
and  turbulence dissipation for plasma heating.
By varying the coronal magnetic field
\citep[based on the 2D, force-free, analytic  
model of][]{1998A&A...337..940B}, they obtain, for a single choice
of {wave amplitude and correlation length at the base}, 
a realistic latitudinal
 range of slow and fast wind conditions, as measured
by the  Ulysses spacecraft \citep[e.g.][]{1992A&AS...92..237B}.
Hence, the characteristics of the solar wind solution along each open
flux tube are solely determined by the properties of the magnetic field,
in agreement with the prescription of empirical models, such WSA
\citep{1990ApJ...355..726W,2004JASTP..66.1295A},
or empirically-driven, such as \citet{2001JGR...10615889R}. 

Using the 1D, time-dependent, solar wind  model of \citet{2014ApJ...784..120L},
we explore an almost identical magnetic configuration, calculating solar wind
solutions along selected field lines,  and obtaining 
results that confirm the results of \citet{2007ApJS..171..520C}.
However, by examining the pressure gradients between neighboring
flux tubes, we suggest that observed latitudinally sharp transitions between  slow and fast flows,
obtained with a single choice of turbulence parameters in the 1D models,
may actually be smoothed out in multidimensional calculations.
 
This paper is organized as follows:
in Sec.~\ref{sec-model}
we present the characteristics of our model, comparing similarities and
differences with that of \citet{2007ApJS..171..520C}.
 In
Sec.~\ref{sec-results} we present  our solutions and show in what sense
they confirm the results of \citet{2007ApJS..171..520C} and in  
what sense they question them.
Our conclusions follow.


\section{MODEL DESCRIPTION}
\label{sec-model}
Our 1D model of the solar wind is based on  the treatment of turbulence
dissipation and wave pressure acceleration presented in 
\citet{2010ApJ...708L.116V}. 
As described in more details in \citet{2014ApJ...784..120L},
the model solves along an open magnetic field line  the following set of time-dependent, 1D HD equations:
\begin{eqnarray}
\frac{\partial \rho}{\partial t}
  &=&  
-  \frac{1}{A}\frac{\partial}{\partial s}\left ( A U \rho \right ) , \label{eq-rho}
  \\
\rho  \frac{\partial U}{\partial t} &=& 
- \rho  U\frac{\partial U}{\partial s} - \frac{\partial }{\partial s}(p+p_w) + g_s \rho
+ \mathsf{R}_s  
 + \frac{1}{s^2}\frac{\partial}{\partial s} 
\left (s^2 \nu  \rho \frac{\partial U}{\partial s} \right ) , \label{eq-mom}
 \\
\frac{\partial T}{\partial t}
 &=& -U\frac{\partial T}{\partial s}
  -(\gamma -1)\left ( T {\frac{1}{A}\frac{\partial}{\partial s}A} U 
 -\frac{m_p}{2 k \rho}
   \left( {\frac{1}{A}\frac{\partial}{\partial s}A} q 
    -  n_en_p{Q(T)}+{H}\right) \right).
\label{eq-T} 
\\
\frac{\partial z_\pm}{\partial t}
&=&
-[U\pm V_a] \frac{\partial z_\pm}{\partial s}+
R_1^\pm z_\pm +R_2^\pm z_\mp - \frac{|z_\mp| z_\pm}{2\lambda_\odot \sqrt{A/A_\odot}
}, \label{eq-zpm}
\\
R_1^\pm&=&-\frac{1}{2}[U\mp V_a]\left ( \frac{\partial \log V_a}{\partial s}
+\frac{\partial \log A}{\partial s} \right ),  \\
R_2^\pm&=&\frac{1}{2}[U\mp V_a]\frac{\partial \log V_a}{\partial s}, \\
H&=&  \rho  \frac{|z_-| z_+^2+ |z_+| z_-^2}{ 4 \lambda_\odot \sqrt{A/A_\odot}}, 
 \\
p&=& 2 n k T \\
p_w&=& \frac{1}{2}\rho \frac{ (z_- -z_+)^2}{8}, \\
\mathsf{R}_s&=& \rho z_+ z_- \frac{\partial \log A}{\partial s}.
\end{eqnarray}
With $s \geq  R_\odot$ we indicate the distance along a  magnetic field line, which is generally different
from the radial coordinate $r$; $p$, $T$,  $U$, and $\rho$, are the plasma pressure, temperature, velocity,
and density. The number density, $n$, is assumed to be equal for protons and electrons.
$k$ is Boltzmann constant.
$g_s=g_0 R_\sun^2 \mathbf{\hat{b}\cdot \hat{r}}/r^2$ is 
the gravitational acceleration parallel to the magnetic field line
($\mathbf{\hat{b}}$).  The  kinematic viscosity is  $\nu$. Along the field line, we call  $A(s)=1/B(s)$
the area factor, which corresponds to  the inverse of the magnetic field magnitude $B(s)$.
The field aligned component of the 
vector divergence of the MHD Reynolds stress, 
$\mathbf{R}=(\delta \mathbf{b}  \delta \mathbf{b}/ 4 \pi - \rho \delta
\mathbf{u}  \delta \mathbf{u} )$, is $R_\mathsf{s}$. $\delta\mathbf{u}$ and  $ \delta\mathbf{b}$ are
 respectively the fluctuations of the 
velocity $\mathbf{u}=U(s) \mathbf{\hat{b}} + \delta \mathbf{u}$ and of the
magnetic field, $\mathbf{B}=B(s) \mathbf{\hat{b}} +\delta  \mathbf{b}$, 
with  $\mathbf{\hat{b}} \cdot \delta \mathbf{b} = 0= \mathbf{\hat{b}} \cdot \delta \mathbf{u}$. 
$ p_w = {\delta \mathbf{b}^2}/{8 \pi}  $ is the
wave pressure.  In  
Eq.~(\ref{eq-T}), the polytropic index is
 $\gamma=5/3$.  The radiation loss function $Q(T)$ is as in \citet{1986ApJ...308..975A}. 
$n_p$ and $n_e$ are respectively the proton and electron 
number
densities (which are equal    for a hydrogen plasma). For the heat flux $q$, according to the radial distance, 
either a
collisional (Spitzer's law) or collisionless form \citep{1978RvGSP..16..689H} is employed.
At a distance of $10R_\odot$ from the Sun, a smooth transition between the two forms  occurs 
\citep{1999PhPl....6.2217M}.
In Eq.~(\ref{eq-zpm}),
the Elsasser variables
$\mathbf{z}_\pm=\delta \mathbf{u}
\mp \delta \mathbf{b}/\sqrt{4 \pi \rho} $  \citep{2001ApJ...548..482D} are advanced. 
$\mathbf{z}_+$  represents an outward propagating perturbation along a radially outward 
 magnetic field line, while $\mathbf{z}_-$ is directed inwardly.
 The 
actual direction of $\mathbf{z}_\pm$ is assumed to be unimportant, provided that  it is in
the plane perpendicular to $\mathbf{\hat{b}}$ and that only low-frequency
perturbations are relevant for the heating and acceleration of the plasma. 
Hence, we treat $z_\pm$ as scalars. The Alfv\'en speed along the field line is
$V_a(s)=B/\sqrt{4 \pi \rho}$.  With 
$R_1^\pm $ and $R_2^\pm$ respectively, we indicate  the WKB and reflection terms, 
which are related to the large scale
gradients. $\lambda_\odot$ is the turbulence correlation scale at the solar surface.
Thus the heating function $H$ \citep{1938RSPSA.164..192D,2004GeoRL..3112803M},
$p_w$ and $\mathrm{R}_s$ 
\citep{2011ApJ...727...84U,2012ApJ...754...40U}
can all be expressed in terms of   $z_\pm$.
 At the lower  boundary, since
the solar wind is subsonic, we are allowed to specify temperature and  density, but the velocity must
be  determined by solving
the 1D gas characteristic equations. Since the upper  boundary is placed
beyond all critical points,  the characteristic equations are used for all variables.
The amplitude of
the outward-propagating (from the Sun) wave is imposed in the $z_\pm$ equations.
\begin{figure}
\includegraphics[width=\textwidth]{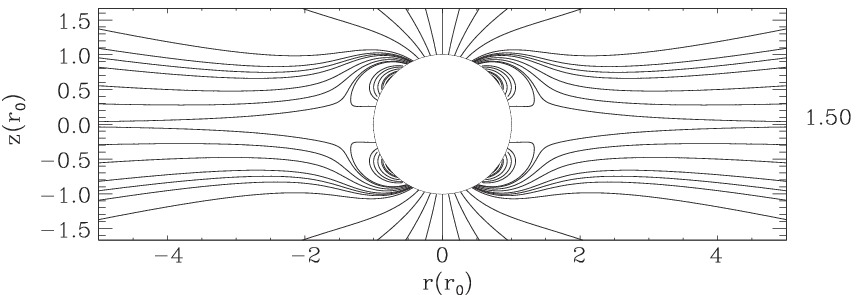}
\caption{Magnetic field in the analytic model of
\citet{1998A&A...337..940B} {for $Q=1.5$}: we solve Eqs.~(\ref{eq-rho}-\ref{eq-zpm}) along
selected field lines. Adapted from Fig.~2 of \citet{1998A&A...337..940B}.
}
\label{fig-banaskiewicz}
\end{figure}

For the present investigation,
we express the area expansion factor $A(s)$ and the gravitational factor
$g(s)$ along selected magnetic field lines calculated
using the analytic model of \citet{1998A&A...337..940B}. This reproduces
the characteristics of the magnetic field of the solar corona and
interplanetary space during solar minimum by combining a dipole, a quadrupole,
and a current sheet (Fig.~\ref{fig-banaskiewicz}). 
The magnetic field becomes
basically radial at large distances from the Sun. 
We use the choice parameters
that makes the last closed field line intersect the Sun at 60 degrees latitude
and $B_r\sim 3.1~\mathrm{nT}$ at 1 AU \citep[$K=1$, $M=1.789$, $a_1=1.538$,
and $Q=1.5$ in Eqs.(1-2) in][]{1998A&A...337..940B}\footnote{Notice that
the third term on the right-hand side of Eq.~(1) of \citet{1998A&A...337..940B}
should be multiplied by $\sgn{z}$}.

The model of \citet{2007ApJS..171..520C} solves a set of
equations  equivalent to
Eqs.~(\ref{eq-rho}-\ref{eq-zpm}) for
mass, momentum, and energy conservation. However, it incorporates
a more sophisticated
treatment of the physics, which includes a number of features 
absent      in our model, 
such  as a                                          
photosphere, treatment of neutral hydrogen, and heating of the chromosphere through 
dissipation of 
a spectrum of sound waves.  Moreover, the
 heating and acceleration of the solar wind through
turbulence dissipation of Alfv\'en waves, which is also present in our
model in the low-frequency limit, 
includes a broad series of frequencies.
The solution methods employed in  the two models are also different: while we
use a time-dependent scheme, \citet{2007ApJS..171..520C} rely on
 an iteration and relaxation method. Since \citeauthor{2007ApJS..171..520C}
set the lower boundary in the photosphere, 
at low heights they modify the
 coronal magnetic field solution
of  \citet{1998A&A...337..940B} 
 according to the model of
\citet{2005ApJS..156..265C}.

\section{RESULTS}
\label{sec-results}
\begin{figure}
\includegraphics[width=\textwidth]{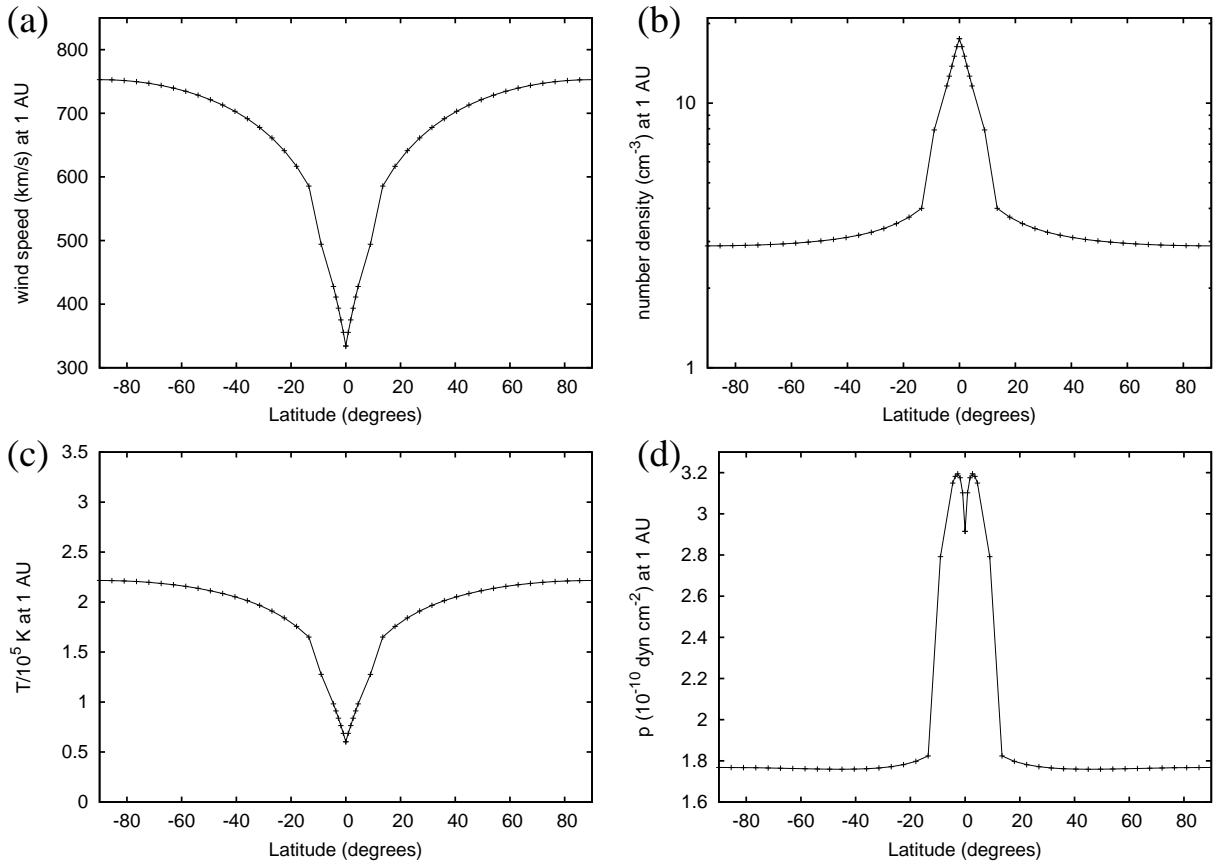}
\caption{Latitudinal dependence at 1AU of 1D solutions that were
 calculated
along field lines of the magnetic field model of  \citet{1998A&A...337..940B} 
traced inward from 1AU:  
(a) wind speed; (b) number density; 
(c) temperature; (d) gas pressure. Each solution is
marked with a cross.
}
\label{fig-1au}
\end{figure}
We solve Eqs.~(\ref{eq-rho}-\ref{eq-zpm})  to obtain  steady-state
 solar wind
solutions along selected field lines of the force-free field of
\citet{1998A&A...337..940B}. In all the solutions we use the same boundary
conditions at the base of the domain, namely $n_0=2\times
10^{12}~\mathrm{cm^{-3}}$ and $T_0=20{,}000~\mathrm{K}$, to
include the transition region and upper chromosphere in the calculation.
A description of these boundary conditions at the base of the chromosphere
is found in  \citet{2005ApJ...621.1098M} and \citet{2013ApJ...773...94M}.
We likewise employ a technique to broaden artificially the
transition region, while maintaining accuracy in
the corona \citep{2009ApJ...690..902L}.
We have used a single combination of values $z_+^\odot=54~\mathrm{km/s} $ and
$\lambda_\odot=0.02~R_\odot$ for Eq.~(\ref{eq-zpm}), which, according to
\citet{2014ApJ...784..120L}, is consistent with a fast stream solution. 
{These values can be compared with those found in
 the model of \citet{2005ApJS..156..265C}.  Since 
the magnitude of the magnetic field of \citet{1998A&A...337..940B} 
varies 
 at the footpoints of open field lines
between
11.8 and 7.7 G from pole to streamer, then, according to Eq.~(51) of 
\citet{2005ApJS..156..265C},  our correlation length $\lambda_\odot$ 
would be of the order of a network flux bundle. Furthermore, 
if we compare Figs.~3 and 11 of 
\citet{2005ApJS..156..265C}, we find a that,
at a height for which their magnetic field strength is the same as in our 
model, their magnitude of the Elsasser variables roughly agrees
with ours, being  between 50 and 60 km/s.
}
A nonuniform mesh is employed with 631 points and $\Delta s$ ranging from
$2.7\times 10^{-4}~R_\odot$ at the solar surface to $7.6~R_\odot$ at 1 AU. 
To damp  unresolved
scales below grid resolution, we add a small kinematic viscosity, such that {the ratio of}  
the associated dissipation
time with the propagation time of Alfv\'en waves  
is $\tau_\nu/\tau_A=5000$  \citep{2009ApJ...690..902L}.  

\subsection{Presentation}
We have first explored the behavior of our model to see whether 
the observed dichotomy at 1 AU  between the fast, rarefied  polar and the slow, denser
equatorial
streams  \citep[e.g.,][]{1992A&AS...92..237B} could
be reproduced with a single choice of turbulence parameters.
Figure~\ref{fig-1au} shows the 
latitudinal dependence at 1AU of wind speed (\ref{fig-1au}a), number 
density (\ref{fig-1au}b), temperature (\ref{fig-1au}c), and pressure
(\ref{fig-1au}d). Each point, marked with a cross, represents a value
obtained with a 1D solution 
calculated
along magnetic field lines traced inward from 1AU. 
We notice that the wind speed transitions from a fast
polar stream to a slow equatorial  as we move towards      
the current sheet. On the other hand,
 the number density of the plasma increases almost one order
of magnitude close to the current sheet. Although the temperature
decreases at lower latitudes, behaving similarly to
the wind speed, the gas pressure turns out 
almost $80\%$ higher at $\pm
15^\circ$ around the current sheet. This result may be compared
with Fig.~12 of \citet{2007ApJS..171..520C}.

\begin{figure}
\resizebox{0.8 \textwidth}{!}{\rotatebox{0}{\includegraphics{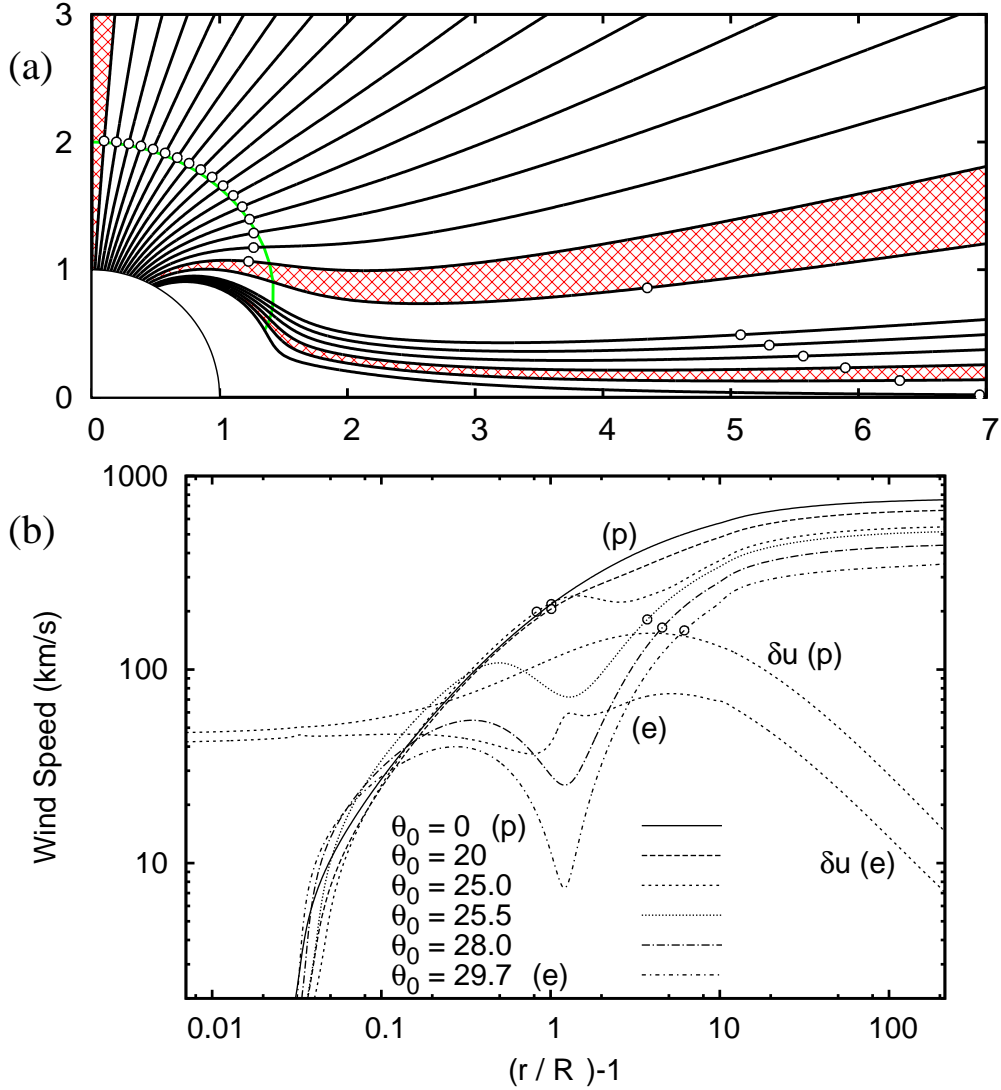}}}
\caption{
(a) Magnetic field lines  from the model
of  \citet{1998A&A...337..940B}. 
The position of the critical points for sound waves for the 1D wind solutions
calculated along them is indicated with circles. 
The green curve is drawn at a distance of 1 $R_\sun$ along of the field lines: 
the plasma pressure along the curve  is plotted in Fig.~\ref{fig-forces}a. 
The red  filling between pairs of
field lines indicates that the acceleration perpendicular to $\mathbf{\hat{b}}$
is plotted in Fig.~\ref{fig-forces}b.
(b)
Wind speed  
 calculated
along several flux tubes prescribed with the same model.
The speed is given as a function of the distance along the field line.
The field lines originate from the solar surface at colatitudes
varying between 0, polar field line, indicated with (p), and 29.7, field line
that  goes close to the equator, indicated with (e). The circles show 
where the wind becomes supersonic. The speeds of the perturbation, $\delta u$,
for the (e) and (p) solutions are also shown.
}
\label{fig-v_dpcranmer}
\end{figure}
Then we have considered how the solar wind speed varies along open magnetic
flux tubes corresponding to the fast and the slow streams.
Figure~\ref{fig-v_dpcranmer}, which can be compared with
Fig.~11 of \citet{2007ApJS..171..520C},
 shows the position of the sonic points and 
the speed of the solar wind calculated
along several magnetic field lines. For each field line, the colatitude of the
foot-point at the solar surface is indicated. The polar field line (p)
originates from the pole; the equatorial field line (e) is the last open field
line, since field lines with higher colatitude are closed.
We  also show the value of the
turbulent velocity perturbation for the  polar and equatorial field lines.
The wind  calculated along 
the polar field line as well as neighboring lines becomes rapidly supersonic
at a length of less than $1~R_\odot$.  Then there is a rapid transition between
25$^\circ$ and $25.5^\circ$ colatitude, with the sonic point moving outwards
and a deceleration zone appearing in its place.               
This  zone becomes increasingly more pronounced as the foot-point
is set at higher colatitudes. In agreement with  what is
 shown in Fig.~\ref{fig-1au}a, the final speed for solutions
computed on magnetic field lines closer
to the equator is considerably smaller.

\begin{figure}
\resizebox{0.8 \textwidth}{!}{\rotatebox{0}{\includegraphics{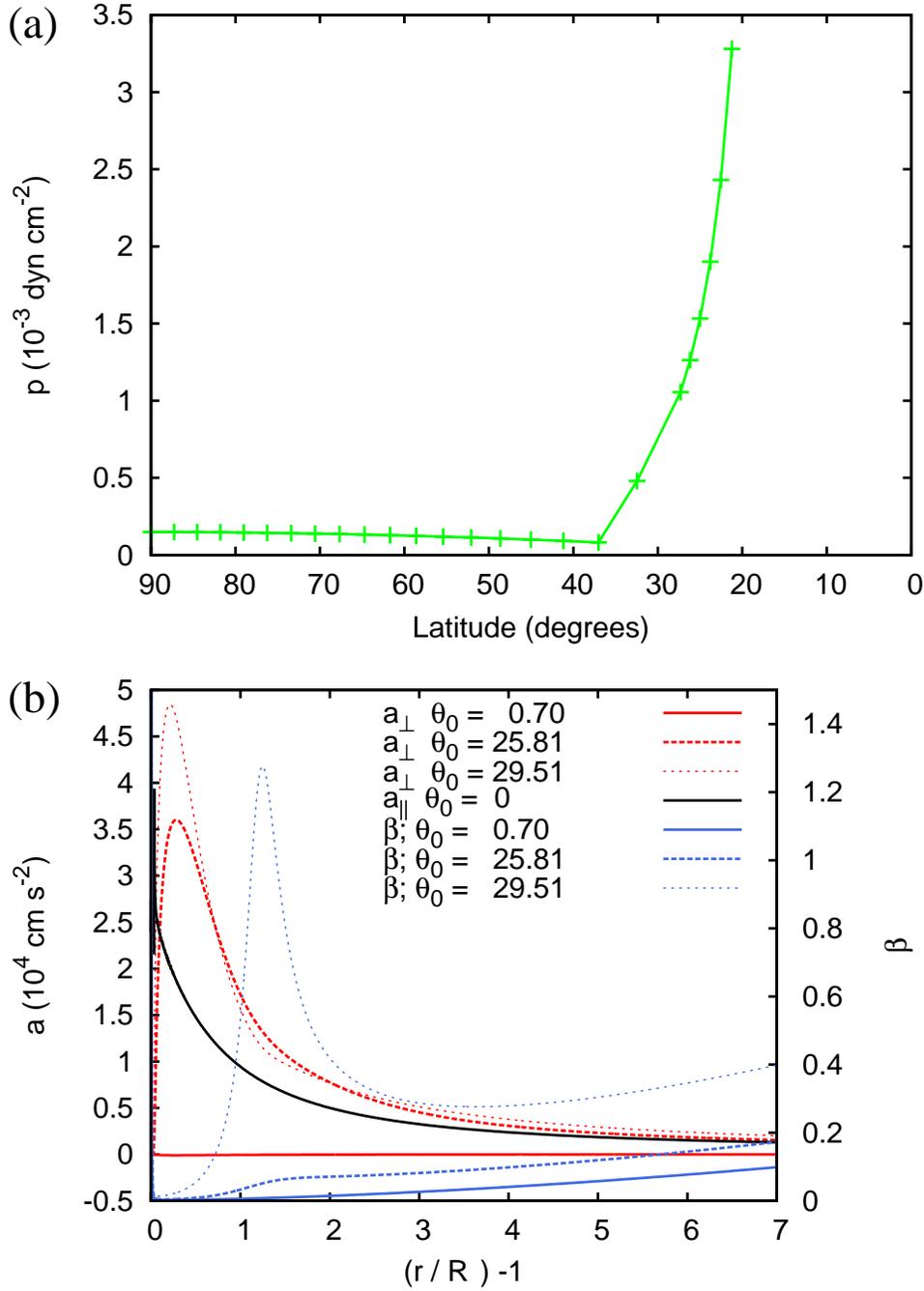}}}
\caption{
(a) The plasma pressure obtained from  the
1D solar wind solutions and evaluated
 along the green curve shown in Fig.~\ref{fig-v_dpcranmer}a.
Each solution  is marked with a cross.
(b) In red: acceleration mainly due to pressure gradients
 perpendicular to $\mathbf{\hat{b}}$, obtained
by differencing the pressure between  two 
1D solar wind solutions, calculated along magnetic 
field lines with colatitudes around $0.70^\circ$, $25.81^\circ$, 
and $29.51^\circ$ respectively. The pairs of
magnetic field lines are shown in red  in Fig.~\ref{fig-v_dpcranmer}a. 
The contribution due to the centrifugal force, although it is included, is minimal. The acceleration is directed towards higher latitudes.
In blue: the corresponding plasma $\beta$ at the same positions. In black: acceleration due to the pressure gradient along the magnetic field for the polar solution.
}
\label{fig-forces}
\end{figure}
Figure~\ref{fig-forces}a shows the pressure obtained with
  the 1D solutions and evaluated along
the green curve of Fig.~\ref{fig-v_dpcranmer}a, which is drawn at
a distance of 1 $R_\sun$ along the field lines. Comparing with 
Fig.~\ref{fig-v_dpcranmer}b, we notice that the pressure increases considerably as we move towards the equator, where,
as predicted by  Bernoulli's principle 
{
(\citealt{bernoulli1738danielis}, p.~231,
or, for a more recent formulation, \citealt{1975hydr.book.....L}, p.~20),}
a velocity stagnation point is present. 

Since both Fig.~\ref{fig-1au} and Fig.~\ref{fig-v_dpcranmer} 
indicate a rapid latitudinal transition between the fast and the 
slow wind, we have calculated the acceleration
perpendicular to the magnetic field, to determine whether
the force-free  magnetic field of \citet{1998A&A...337..940B} 
would be capable of 
accommodating such gradients
without considerable deviation from the analytic form. We 
selected three pairs of magnetic field lines, indicated with a red pattern 
in Fig.~\ref{fig-v_dpcranmer}a, and for each calculated the quantity
\begin{equation}
a_\perp =\frac{1}{\rho}\frac{\Delta p}{\Delta l}+ \frac{U^2}{R_c},
\label{eq-acc}
\end{equation}
where $\Delta l$ is the distance perpendicular to the magnetic field
and $R_c$ is the local radius of curvature. In the present cases, 
the first term on the right hand
side of Eq.~(\ref{eq-acc}) dominates everywhere over the second.
  For the three pairs
of magnetic field lines, we plot $a_\perp$ 
in red in
Fig.~\ref{fig-forces}b. The acceleration is directed  from the current 
sheet towards the North Pole. The magnitudes of the perpendicular acceleration 
are not negligible, since they are
 comparable 
with the values of the acceleration $a_\parallel$ due to the parallel pressure gradient 
computed  along the polar 
field line, which is shown in black in Fig.~\ref{fig-forces}b.
For the three examples, we also show in blue the plasma 
$\beta=4\pi p/B^2 $.  While $\beta$ is very small at high latitudes,
it increases considerably as we move towards the current sheet, close to which  it becomes larger than one.  
It follows that
for field lines with footpoints at larger colatitudes a strong pressure
gradient is expected to build. 
{There is evidence in the literature that this would cause
a restructuring of the magnetic field. For example, in the pioneering work of 
\citet{1971SoPh...18..258P},
an isothermal  model of the corona is calculated 
by imposing the equilibrium of forces normal to the magnetic field.
More recently, \citet{2003ApJ...598.1361V} used an
iterative, 2D model of the solar corona with anisotropic
gas pressure and  found that  the build up of perpendicular
pressure gradients causes a reconfiguration of the magnetic field. Moreover,
the presence of a stagnation point near the streamer cusp, in combination with
the effects of plasma $\beta$  being of order 1 at the cusp, was studied by 
\citet{2002ApJ...565.1275S} and \citet{2005ApJ...624..378N}. }
  Hence, in order to balance these normal
gradients in a self-consistent MHD calculation,
  the equilibrium magnetic field must change. Since 
large variations of the equilibrium field are necessary
to balance the pressure gradients, the solar wind solutions are also
expected to change.
\subsection{Discussion}
Our results and  those of 
 \citet{2007ApJS..171..520C}, in spite of 
quantitative differences,
 are in general qualitative agreement.
The fact that 
our choice of turbulent parameters is different should  not be surprising.
For \citet{2007ApJS..171..520C} include the photosphere
in their model and specify their boundary values at a lower height than
we do.
If we compare
Fig.~\ref{fig-1au}  with Fig.~12 of \citet{2007ApJS..171..520C}, we
notice that our curves for wind speed and temperature
are roughly in between the two sets of models calculated by
\citeauthor{2007ApJS..171..520C}.  However, our density appears 
to be larger in the equatorial region than the density in either sets.
For this reason, although the dependence of
pressure on latitude is not provided in \citet{2007ApJS..171..520C},
the pressure jump as one moves closer to the current sheet has
the opposite sign than in our calculations. Nevertheless, as one moves closer
to
the current sheet,
rough estimates yield for the ``Durham'' set  a pressure jump
$\Delta p \approx 1.5\times 10^{-10}~\mathrm{dyn~cm}^{-2} $, and $\Delta p
\approx 10^{-10}~\mathrm{dyn~cm}^-2 $ for the standard set.  Considering
the high beta ($4\pi p/B^2 \gtrapprox 30$) of the plasma  in that region, these pressure
 gradients are  not likely to appear in a {steady}
self-consistent multidimensional MHD model, since the magnetic field would not
be able to balance them.

From a comparison of Fig.~\ref{fig-v_dpcranmer} with Fig.~11 of
\citet{2007ApJS..171..520C}, we notice that the transition between solutions
with sonic points in the lower corona and those with  sonic points in the
higher corona is at close but not identical latitudes. 
Considering the differences
between the models, we do not find this discrepancy surprising. 
The two figures are  also clearly different close to the solar surface, where
\citet{2007ApJS..171..520C} are able to resolve the photosphere, which is not
included in our model.
 However, 
the main, qualitative characteristics of the solutions appear to be
the same: namely, a faster, high-latitude stream vs.\ a slower stream at
low latitudes. Moreover, the slow-wind solutions present a stagnation point
in either model.

Since 
the pressure differences between neighboring field lines is not investigated in 
\citet{2007ApJS..171..520C}, it is difficult to ascertain the formation
of strong gradients.
However, considering that their solutions exhibit so many characteristics
 similar to ours, 
we surmise that non-negligible
perpendicular pressure gradients are probably present also in the model
of \citeauthor{2007ApJS..171..520C}. In a self-consistent 2D MHD model with
turbulence dissipation,
the  magnetic field structure would adjust to counterbalance these pressure
gradients. As a  consequence of that, changes in the magnetic field would also
modify the speed, density, and temperature of the solar wind. The position of
the sonic points  and the latitudinal bifurcation of slow and fast streams
would also necessarily be modified.

\section{Conclusions}
We have revisited the results of 
\citet{2007ApJS..171..520C} 
using our
1D, time-dependent model of the solar wind of,
which incorporates turbulent dissipation of
Alfv\'en waves  to provide
heating and acceleration of the plasma.
We have obtained solar wind solutions along selected magnetic field lines
of the 2D, analytic model of \citet{1998A&A...337..940B}.
In spite of the unequal level of sophistication of the models, we can confirm
 the main conclusions of \citet{2007ApJS..171..520C}, namely that a single
choice of turbulent parameters specified on the
solar surface is sufficient to provide solutions reproducing the fast and slow
streams of the solar wind and the rapid latitudinal transition
between the two regimes, as in situ measurements show us.
Thus
the characteristics of  solution along each flux tube appear to
be  dictated simply by the properties of the magnetic field line, namely
expansion factor, magnitude of $B$, and inclination from the radial direction.
 However, it is
not surprising that our choice of turbulent parameters at the base
of the computational domain is not the same as
that of \citet{2007ApJS..171..520C}, since we do not include the photosphere
in our calculation as they do.

Then we have investigated the presence of 
perpendicular pressure gradients
between neighboring field lines. We have found evidence for such gradients in
our solutions and we have argued that they likely appear  also in the
results of \citet{2007ApJS..171..520C}. Therefore, we believe that
the configuration with strong latitudinal gradients separating the heliosphere
in slow wind and fast wind sectors, if used as initial condition
in a self-consistent 2D MHD simulation, will be out of equilibrium and
rapidly evolve into something different. It is likely that the latitudinal
 discontinuities
both in the distance of sonic points and in the properties of the winds
will be smoothed. That being the case, a different choice of turbulence
parameters (e.g., as a function of the latitude of the foot points of
the magnetic field lines) will be necessary to reproduce the Ulysses 
measurements or the predictions of (semi)-empirical models. 
It is therefore paramount to perform multi-dimensional
MHD simulations
to verify whether we are right on this point. 
We plan to perform such simulation with our multi-dimensional MHD model, in which turbulent dissipation 
will be implemented according to 
 the present, relatively simple yet accurate, formulation.

Needless to say, our conclusions do not imply that 1D models should be cast away. 
On the contrary, the usefulness of 1D models remains strong: it will always be easier
to incorporate and study new physical effects at least initially in 1D wind models; also, they remain invaluable tools for explorative parameter
studies; finally, they represent the benchmark on which multi-dimensional
models must be tested on. However, the constraints underlying 1D models, and in particular the
assigned geometry of the flow, means that development of multi-dimensional models remains fundamental and should be carried out synergistically 
with more sophisticated 1D models.

\acknowledgements{
This work was supported by AFOSR,  NASA's LWS Slow Wind TR\&T, HTP, and
strategic capabilities programs.
We are grateful for the use of 
computing resources at NASA Pleiades  and NSF Stampede.
MV was supported by the NASA Solar Probe Plus
Observatory Scientist contract. 
}

\bibliography{mybib}
\end{document}